# The Effect of Externally Applied 3D Fields on NSTX Edge Turbulence

Michael Hanson[1,2], Dmitri Orlov[1], Stewart Zweben[3], Andreas Wingen[4]

[1]University of California San Diego
[2]Commonwealth Fusion Systems
[3]Princeton Plasma Physics Laboratory
[4]Oak Ridge National Laboratory
(Dated: October 2022)

# Abstract

We report on a study of the structure of edge turbulence in NSTX H-mode discharges with applied n=1 and n=3 non-axisymmetric magnetic perturbations. The edge turbulence is diagnosed in NSTX using the gas puff imaging (GPI) system to understand how these 3D fields affect edge transport. The presented database study uses a selection of discharges that have a change in the RWM coil current during the GPI gas puff. We compare the turbulence before and after this change in the applied non-axisymmetric fields, and compares the turbulence between the two states. Our findings show that these 3D fields do not have a strong, statistically significant impact on the average poloidal size, autocorrelation time, or relative fluctuation levels of the turbulence. In addition, we find that the displacement of the radial location of the GPI signal peak intensity is correlated with the shift in the perturbed separatrix location as predicted by the 3D vacuum magnetic field code MAFOT. We find that in general the non-axisymmetric fields from the RWM coils locally displace the edge flux surface as expected even when the self-consistent plasma response is not included, and that the model reproduces the localized shift observed by GPI.

## I. Introduction

When a magnetically confined plasma is heated above a threshold heating power level is exceeded, it can transition from a low confinement (L) mode to a high confinement (H) mode. L-mode plasmas generally are turbulent and have poor energy confinement properties. In contrast, H-mode plasmas are characterized by an energy confinement time increase of two or more times times over their L-mode counterpart. Due to the reduced levels of transport in the H-mode, there exist steep pressure gradients at the plasma edge. Large pressure gradients in the H-mode periodically release large bursts of heat and particles, edge localized modes (ELMs), onto the reactor wall which can be damaging. While these ELMs are of a manageable size in current experimental devices, when scaled up to an ITER-like machine, they will be large enough to damage the wall and divertor [Loarte 2014]. Because operating in H-mode is essential for ITER to achieve its mission goals with a fusion gain (Q=fusion power/input power) of 10, ELM control for ITER is essential as well.

Externally applied 3D magnetic fields and more specifically, resonant magnetic perturbations (RMPs), are one way to control these ELMs [Evans 2004]. Magnetic perturbations are a small

($\delta B/B_T \sim 10^{-4}$) disturbance of the magnetic field and when they have the same helicity as a target rational magnetic surface, they are said to be resonant. There are many sources of these 3D fields in tokamaks, both intentional and intrinsic. As seen in figure 1, NSTX has 6 ex-vessel coils that wrap 360 degrees around the midplane of the device. These coils, referred to as Resistive Wall Mode (RWM) coils, are one source of these fields and were initially installed to control the resistive wall mode instability [Sontag 2007] before being used for ELM mitigation experiments [Canik 2010].

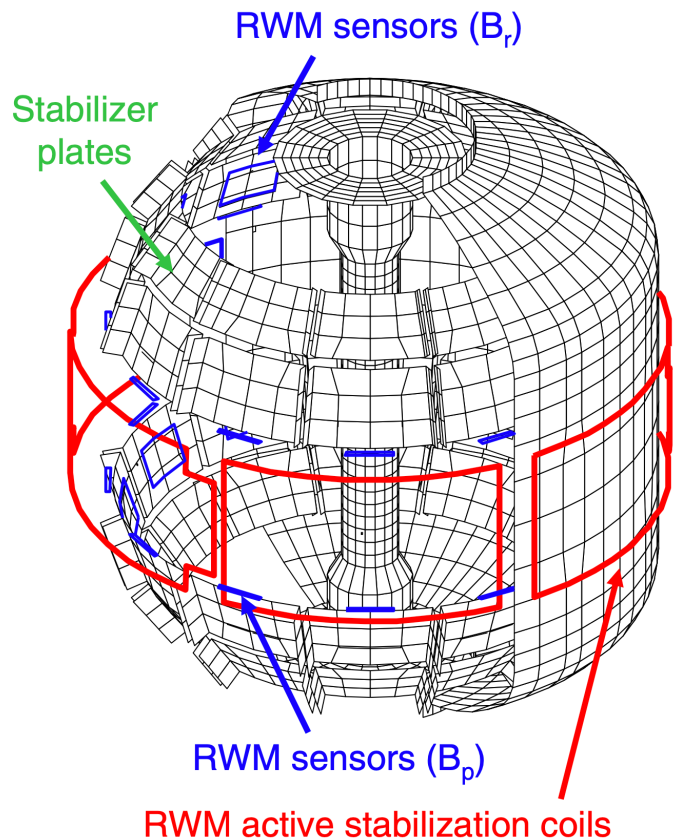

**Figure 1.** NSTX with the RWM coils shown in red [Sabbagh 2010].

The process through which particles and energy flow out of the main plasma and are lost through the separatrix is referred to as edge transport. There are several mechanisms that cause this edge transport, ELMs included, but most are complicated and involve random fluctuations of either the local electric or magnetic field. Experimental validation of transport theory is an important step in better understanding this complex phenomena.

The Gas Puff Imaging (GPI) diagnostic [Zweben 2014] is used to image edge turbulence structures in tokamaks via a puffed cloud of neutral gas that interacts with the electrons in the plasma. This gas, typically deuterium in NSTX, can penetrate several centimeters into the separatrix. The electrons of the plasma excite the electrons of the neutral gas to a higher energy level and when these electrons transition to a lower energy level, light is emitted. Of this light, one of the brightest lines emitted corresponds to Balmer-alpha photons with a wavelength of 656.3 nm.

Light of this wavelength is in the visible part of the spectrum and can be imaged using a camera with an appropriate filter. The intensity of this light is a function of the local neutral atom density (from the puff), local electron temperature, and local electron density.

Fluctuations of GPI intensity can be interpreted as fluctuations in either the local electron density or temperature but have been shown to be largely a function of electron density in the tokamak-edge relevant temperature regimes [Zweben 2017]. The Phantom 710 camera used in NSTX experiments has a frame rate of ~400,000 frames per second with each frame cvering 24 cm in the radial direction and 30 cm in the poloidal. This gives a temporal resolution of about 2.5 microseconds and a spatial resolution of ~1cm (64x80 pixels). A photo of the NSTX GPI hardware is seen in Figure 2. For NSTX, the image is centered at (R,Z) = (149.1, 19.9) cm and at a toroidal angle of 62 degrees [Zweben 2014].

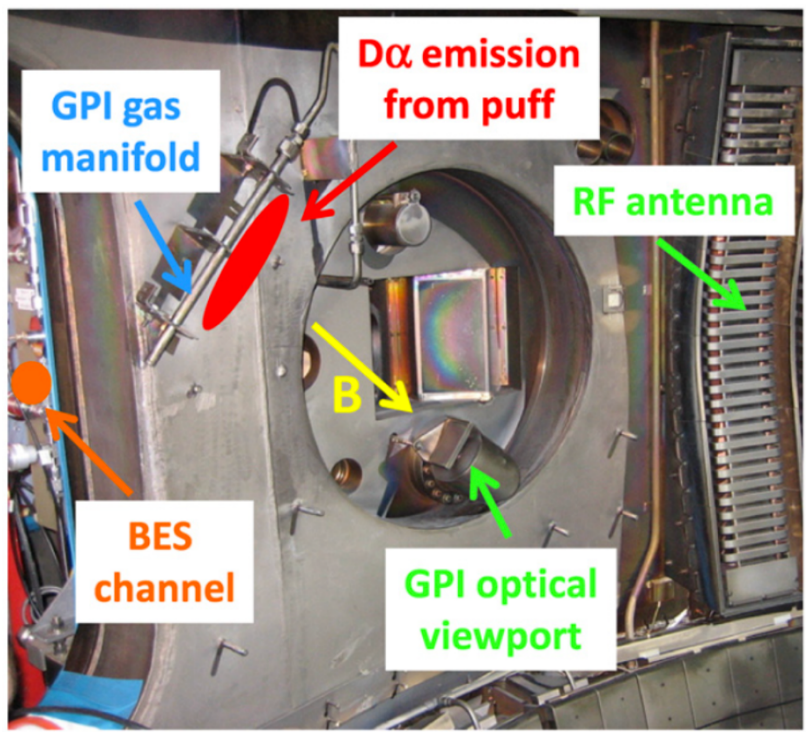

**Figure 2.** GPI hardware in NSTX for the 2010 data [Zweben 2014].

Analyzing the plasma response to externally applied 3D fields is crucial to understanding how these fields affect edge transport in modern devices. One possible transport effect of the 3D fields would be due to a stochastic magnetic field that is not varying in time, just in space. Because GPI measures only the time dependent temperature and density fluctuations, it cannot directly measure the stochastic magnetic fields. However, the 3D fields may indirectly affect the edge fluctuations, which can then be measured with the GPI diagnostic.

Experiments in the past have looked at the effect of 3D fields on edge transport in tokamaks using Beam Emission Spectroscopy [Kriete 2020, Williams 2020, McKee 2013], and probes [Vianello 2015], as well as other techniques. In general, these studies have found that 3D fields result in an increase in density fluctuations in L-mode cases, and most significantly in the edge region of 0.6 < ρ < 0.95 [Williams 2020]. It is also observed that in L-mode plasmas, poloidal velocities are decreased with the application of 3D fields, with RMPs causing a more significant decrease [Kriete 2020]. In the present study, we measure the change in edge turbulence in H-mode plasmas as a result of externally applied 3D fields in the NSTX tokamak with the GPI diagnostic.

## II. Methods

### a. Database

The database for this study consists of a 21-discharge subset of the NSTX 2010 GPI database of 322 shots [NSTX2010]. The 21 H-mode discharges used for this study have a simultaneous GPI data and variation in the RWM coil current and so by extension, a change in the externally applied 3D field. A more detailed description of these variations in the application of the 3D field are given in Table 1. These shots have been chosen such that the fluctuations in the GPI signal are not dominated by MHD fluctuations (~10's of kHz). In these discharges, the RWM coil current amplitude is constant before and after the change in the application of the fields, and plasma parameters are generally constant ($B_T$, $I_P$, shape) over the time of interest. 15 of these discharges have n=3 perturbation fields applied, while 6 shots have an n=1 perturbation rotated 360 degrees toroidally around NSTX [Kimin 2015] providing an opportunity to look for any toroidal phase dependencies.

**Table 1.** Different classifications of changing 3D fields included in this study.

| Type of RWM current change | Mode number | Number of shots in database |
|---|---|---|
| On → Off | n=3 | 3 |
| Off → On | n=3 | 6 |
| Polarity Switch | n=3 | 6 |
| Polarity Switch | n=1 | 6 |

### b. Turbulence Analysis

The GPI data for this study is stored as a series of frames with recorded intensity counts for each of the 64 radial and 80 vertical pixels. Figure 3 shows a single frame GPI image from one of the discharges in our database. The location of the separatrix (as determined by MSE-constrained magnetic equilibrium reconstruction in EFIT) as well as the shadow from the RF limiter are overlayed.

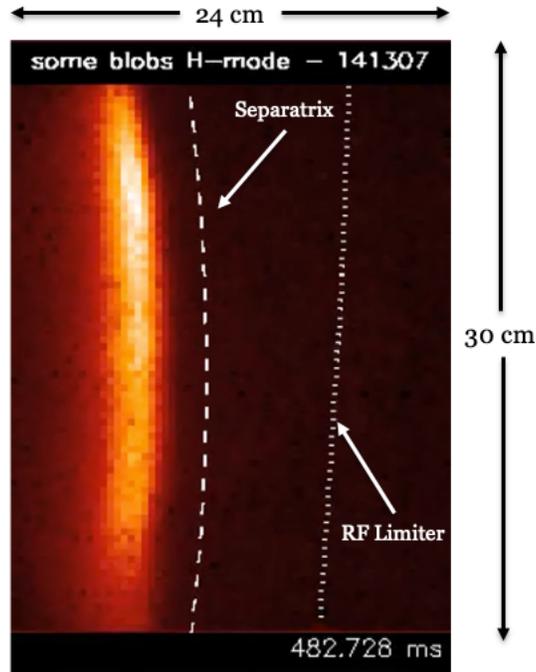

**Figure 3.** Example of GPI data from the 2010 NSTX library [NSTX2010].

The time window of interest is 20ms long and spans from 5ms before to 15ms after the change in the application of the 3D fields. Figure 4 shows the 3 temporal regimes in the 20ms time window used in this work to analyze different cases of 3D field applications: before, during, and after. The first phase, $t_{start}$ → $t_{change}$, represents the "before" period with $t_{start}$ being ~5ms before the change in RWM coil current and $t_{change}$ being the time when the current changes. Whether this "before" period has applied 3D fields depends on the specific shot but both the "before" and the "after" states are considered steady state with respect to the varying 3D field. The middle phase, $t_{change}$ → $t_{change}$ +10ms, corresponds to the transient phase of the applied field.

The internal magnetic field sensors (shown in Figure 1) indicate an average 8-10ms delay exists after a change in the coil current before the full effect of the externally applied 3D fields is measured inside the vacuum vessel and by extension, felt by the plasma [Canik 2010]. The same relationship is seen in Figure 4 using one of the shots from the database. The final window, $t_{change}$ +10ms → $t_{change}$ +15ms, is the period where it is assumed that the effects of the changed 3D field are fully felt by the plasma. We recognize that other changes may be happening in the plasma over a 20ms time window, but we mitigate this by selecting shots where the heating power, plasma current, and general plasma geometry stays constant over this period.

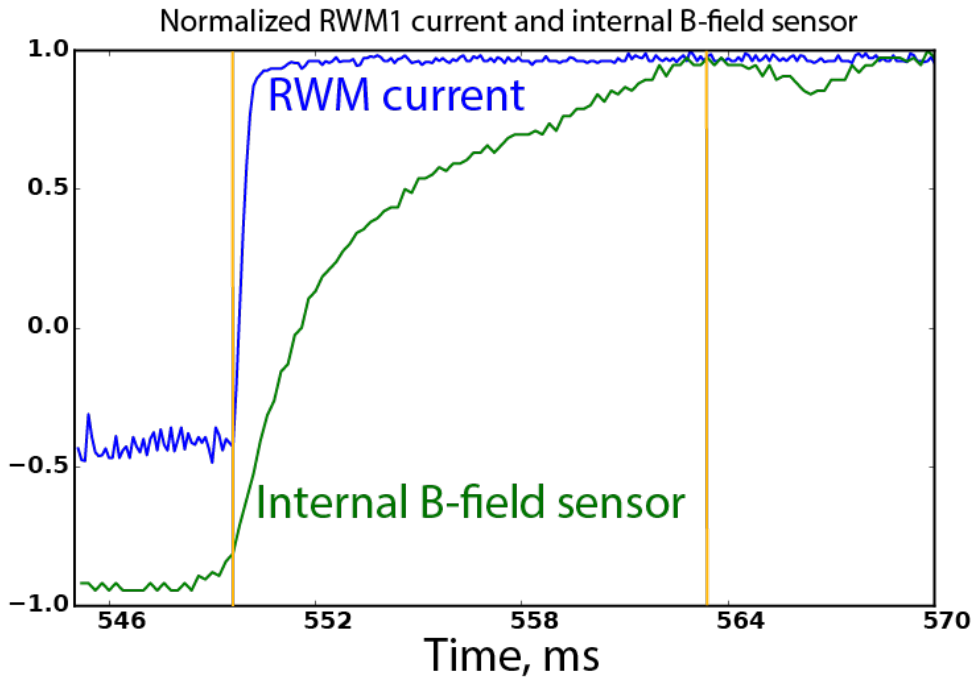

**Figure 4.** Time series of the normalized RWM coil current in black and normalized $B_r$ sensor signal in red showing the 8-10ms penetration time for the externally applied magnetic field.

For each shot, turbulence quantities are calculated and averaged over a specified area in the GPI image and then recorded as functions of time. The first of the areas chosen is a box with a radial span that starts 3cm inside the separatrix and extends to the separatrix, and covers 15cm in the vertical direction, centered on the midplane of the camera view. The second box spans from the separatrix to 3cm outside of the separatrix and has the same vertical dimension as the previous box. Separating the analysis into these two boxes allows for a quick comparison inside and outside of the separatrix.

In each of the two boxes, and for each of the three time periods of interest, three quantities are calculated allowing the turbulence to be characterized and compared. The first is the relative fluctuation level, which describes the magnitude of the GPI intensity fluctuations. The second quantity is the autocorrelation time, or a measure of how long the signal at a point in the image stays the same and is inversely proportional to the average fluctuation frequency. The third quantity is the correlation length and is taken to be the average poloidal size of the turbulence structures.

An example of this analysis is shown in Figure 5. For each pixel in the specified area, and for each time step, a time series of its intensity over 0.5ms around this time is examined. The standard deviation and mean of this series is recorded for each pixel and then averaged over the frame. The vertical yellow bars serve to separate the 3 time phases mentioned above.

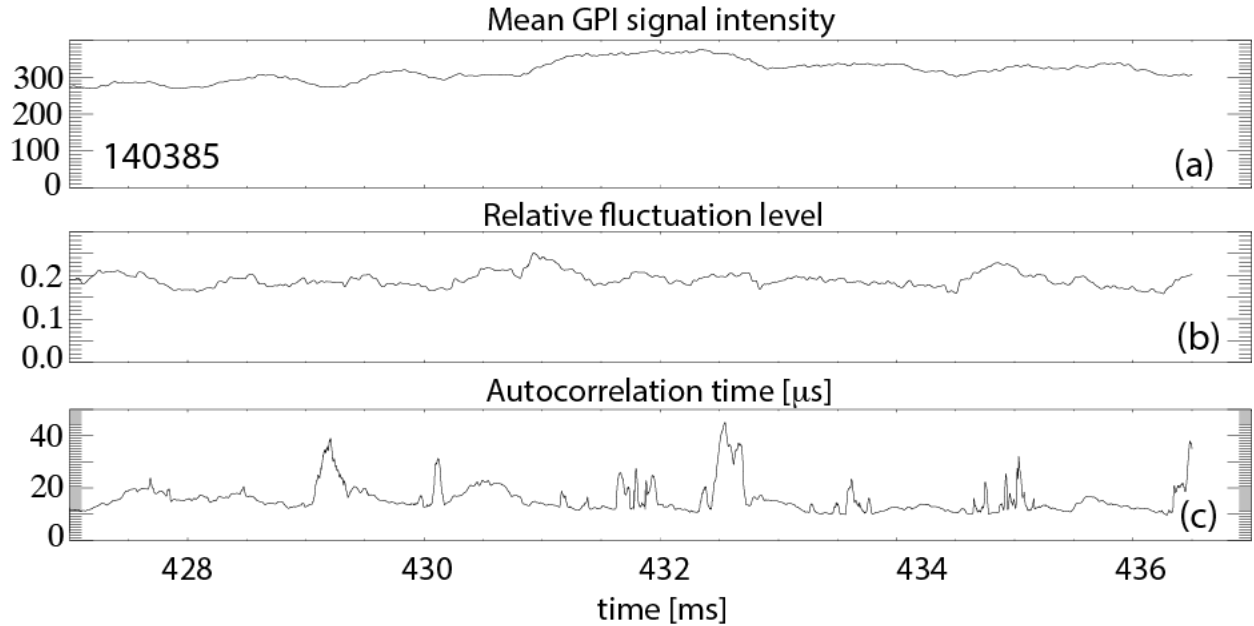

**Figure 5.** (a) Mean intensity of a box inside of the separatrix. (b) Relative fluctuation levels inside of this box. (c) Autocorrelation time of the same box.

For the purposes of this analysis, the relative fluctuation level, shown in the middle plot of Figure 5, is the standard deviation of the intensity divided by its mean. The autocorrelation time is calculated by looking at the autocorrelation function of the same 0.5ms time series described above with a lag that ranges from 1 to 40 frames. With each frame being 2.5 microseconds, the autocorrelation time is taken as the frame in the autocorrelation function that is closest to 0.5, i.e. half width, half max. The three values are calculated for each pixel in the region of interest, and for each time step, and then averaged over said region. That is, the mean of the mean over a region, the mean of the standard deviation over a region, and the mean autocorrelation time over a region.

Figure 6 shows a more detailed spatial analysis that is done by calculating the radial profiles of the average intensity, relative fluctuation levels, and autocorrelation times before and after the change in coil current. In Figure 6, the solid lines are the profiles before the change, and the dashed lines are the profiles afterwards.

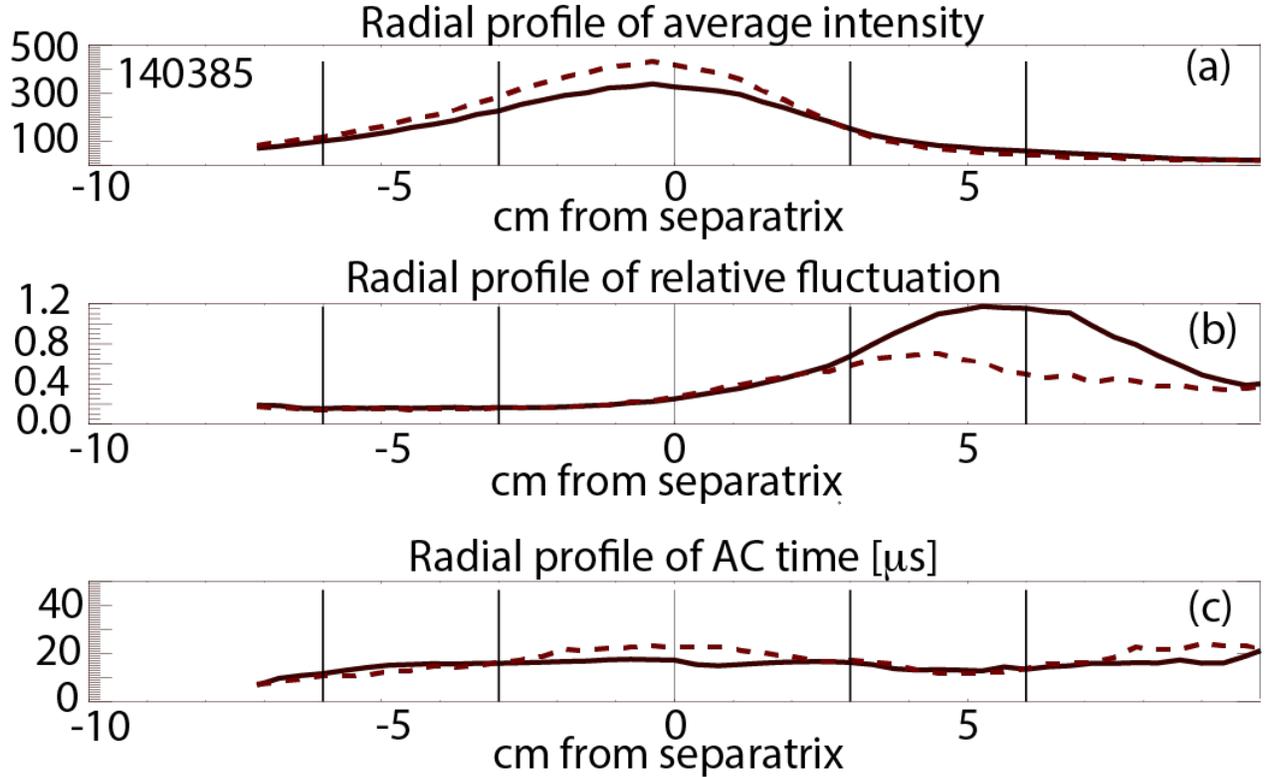

**Figure 6.** Radial profiles of the normalized average intensity (a), relative fluctuation (b), and autocorrelation time (c) for the period before (t=446 ms) and 15ms after (t=461 ms) a change in the RWM coil current in NSTX discharge 140385.

The average poloidal size of turbulence structures in the camera view is calculated by finding the cross-correlation length with zero-time lag [Zweben 2017]. Starting at the pixel corresponding to a point 6cm inside the separatrix and at the midplane of the image, the correlation between this pixel and one displaced vertically is found. The full width, half maximum convention is used to determine the correlation length, which then, by using the conversion from vertical pixel to cm (3/8 $\frac{cm}{pixel}$) we can determine the average poloidal size at reach radial point. This size is then averaged over the entire radial span, and then averaged over 1ms.

### III. Results and Discussion

#### a. Turbulence Analysis

##### i. 3D Fields On → Off

For the three cases where the RWM coil current switches from on to off during the GPI puff, Table 2 shows the average size, autocorrelation time, mean intensity, and relative fluctuation levels for before and after the switch.

**Table 2.** Summary of results for the 3 on-off switch cases

|                                              | Before      | After       |
|----------------------------------------------|-------------|-------------|
| Relative Fluctuation level (Inside Separatrix)  | 0.21 ±0.02  | 0.26 ±0.01  |
| Relative Fluctuation level (Outside Separatrix) | 0.41 ±0.04  | 0.53 ±0.02  |
| Autocorrelation Time [µs] (Inside Separatrix)   | 10 ±1.25    | 9 ±1.25     |
| Autocorrelation Time [µs] (Outside Separatrix)  | 11 ±1.25    | 13 ±1.25    |
| Average Poloidal Size [cm]                      | 3.7 ±0.2    | 4.2 ± 0.7   |

While no significant systematic change is seen in either the autocorrelation time or average size, a weak trend is observed in the relative fluctuation level. As shown in Figure 7 and Table 2, for both inside and outside of the separatrix, an increase in the relative fluctuation is seen when the 3D fields are turned off. This trend is observed in both the radial profiles and the time evolution of these values. Because there is no interpolation between frames, an uncertainty of ±1.25 microseconds exists, as shown in the error bars of the autocorrelation time plots. The error bars on the average size and relative fluctuation level plots represent one standard deviation above and below. It is recognized that because this data set consists of only three shots, the standard deviation is not statistically meaningful.

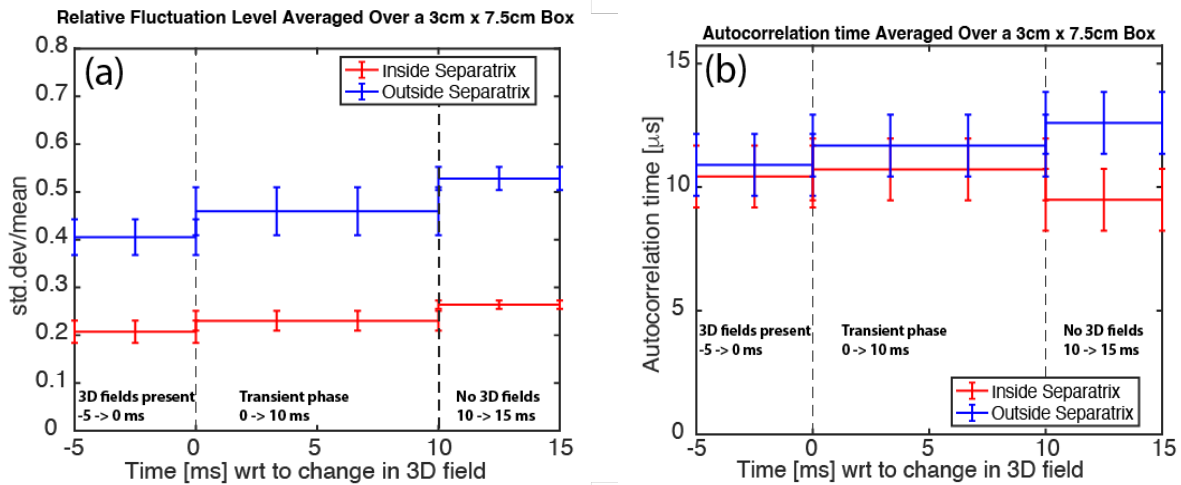

**Figure 7.** Relative fluctuation levels (left) and autocorrelation times (right) for inside (red) and outside (blue) the separatrix. At t=0, the RWM current drops to zero from ~500A.

### ii. 3D Fields Off→ On

Figure 8 shows the autocorrelation time and relative fluctuation levels for the six cases where the RWM coil current turns on during the GPI puff and Table 3 summarizes all results for these cases. Again, no significant systematic change is seen in either the autocorrelation time or average size. However, a weak trend is observed in the relative fluctuation level that is consistent with the previous on-to-off case.

Table 3. Summary of results for the 6 off-on switch cases

|  | Before | After |
|---|---|---|
| Relative Fluctuation level (Inside Separatrix) | 0.25 ±0.05 | 0.18 ±0.06 |
| Relative Fluctuation level (Outside Separatrix) | 0.4 ±0.1 | 0.4 ±0.2 |
| Autocorrelation Time [µs] (Inside Separatrix) | 12 ±1 | 15 ±1 |
| Autocorrelation Time [µs] (Outside Separatrix) | 13 ±1 | 14 ±1 |
| Average Poloidal Size [cm] | 4.3 ±0.6 | 6±2 |

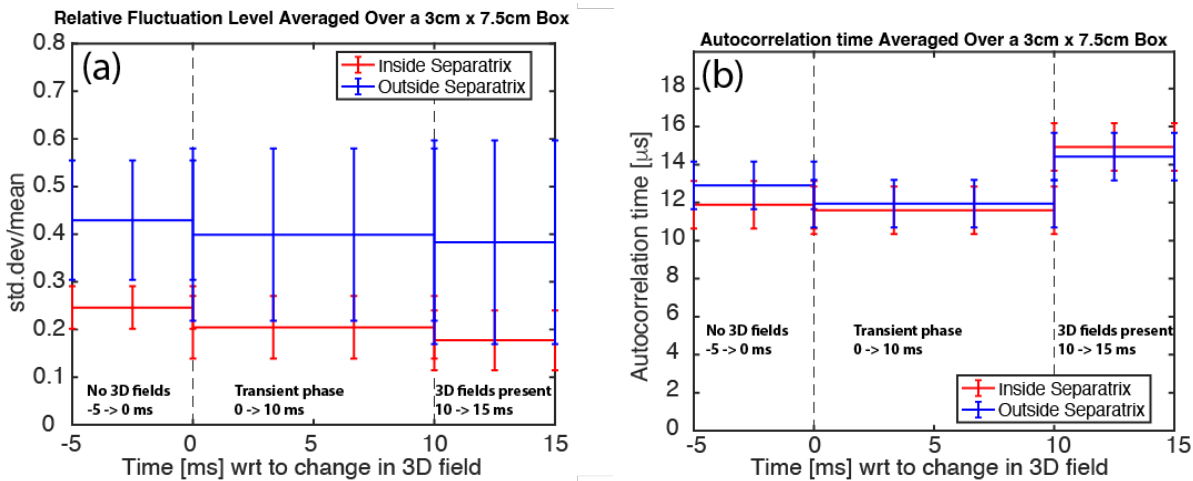

**Figure 8.** Relative fluctuation levels (left) and autocorrelation times (right) for inside (red) and outside (blue) the separatrix. At t=0, the RWM current increases from zero to ~500A.

When the current is turned on and the 3D field applied, a decrease in the relative fluctuation is seen both inside and outside the separatrix. This trend is observed in both the radial profiles as well as the time evolution of these values. This trend is less convincing than the on to off cases and has most of the error bars overlapping.

### iii. Polarity Switch of the 3D Fields

The final subset of shots in our database are cases where the polarity of the coil is switched during the GPI puff. For these cases, a typical current trace is shown in Figure 9. Here, the "before" period represents a time when the RWM coils are being used for n=3 error field correction, while afterwards, a n=3 perturbation is applied with a larger current but with the opposite polarity [Ahn 2017].

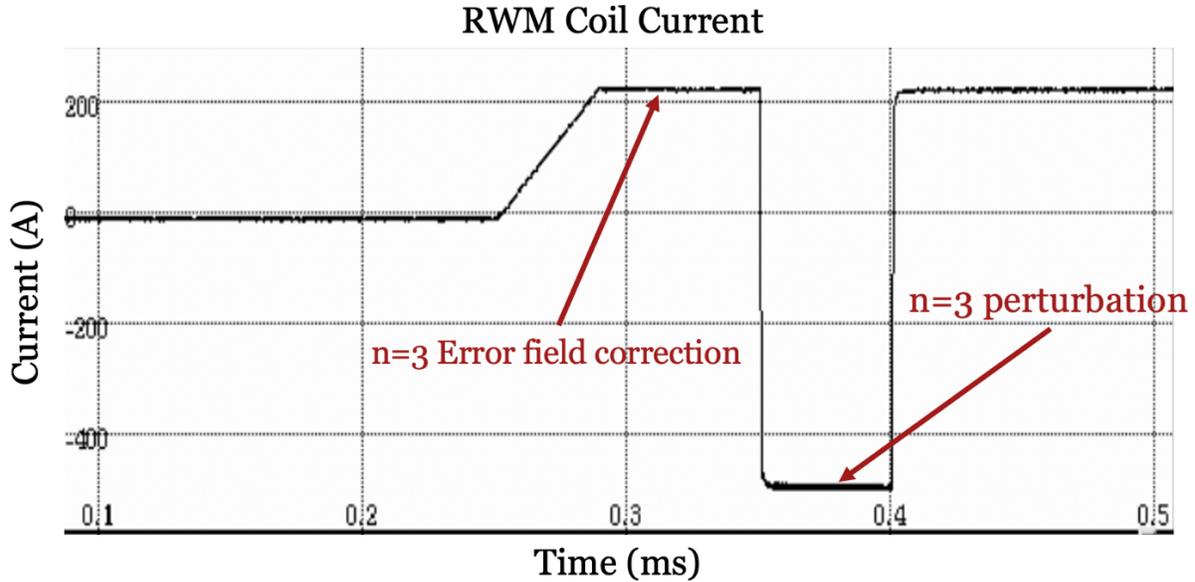

**Figure 9.** Typical RWM current traces for the shots where the coil polarity is switched during the GPI puff.

While the same three metrics are compared, the comparisons for this subset are made more difficult by the variation between shots. Whereas the previous two subsets all had comparable coil current magnitudes, around 500A, this subset has a wide range of current magnitudes, ranging from 400 to 1000A. Figure 10 shows the data for relative fluctuation levels as well as the autocorrelation time.

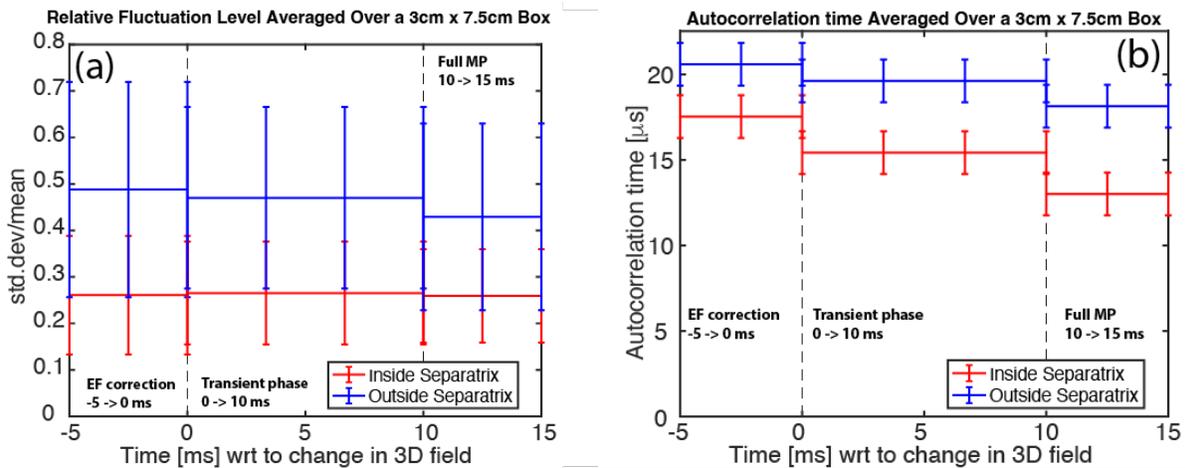

**Figure 10.** Relative fluctuation levels (left) and autocorrelation times (right) for inside (red) and outside (blue) the separatrix. At t=0, the RWM coils go from applying error field (EF) correction to deliberate magnetic perturbations (MPs).

This set shows no statistically significant trends in the autocorrelation time, relative fluctuation levels, or the average poloidal size, nor were there seen any systematic change in the radial profiles of these values.

Table 4. Summary of results for the 6 n=3 polarity switch cases

|  | Before | After |
|---|---|---|
| Relative Fluctuation level (Inside Separatrix) | 0.3 ±0.1 | 0.3 ±0.1 |
| Relative Fluctuation level (Outside Separatrix) | 0.49 ±0.2 | 0.4 ±0.2 |
| Autocorrelation Time [μs] (Inside Separatrix) | 18 ±1.25 | 13 ±1.25 |
| Autocorrelation Time [μs] (Outside Separatrix) | 21 ±1.25 | 18 ±1.25 |
| Average Poloidal Size [cm] | 5.9 ±0.7 | 4.6 ±0.5 |

### b. Separatrix displacement by the 3D fields

The radial location of the peak intensity of the GPI signal (shown in Figure 6a) can be used to estimate the location of the plasma separatrix. The shift in the radial location of the separatrix can be observed and measured during the changes in the application of the 3D fields. This is due to the splitting of the perturbed separatrix and the formation of the non-axisymmetric lobes on the high and low field sides of the tokamak when the 3D perturbation fields are turned on. Respectively, the shift in the location of the separatrix is due to the disappearance of the non-axisymmetric lobes when the perturbation fields are turned off. The changes in the toroidal phase of the perturbation fields typically result in the toroidal shift of the non-axisymmetric separatrix surface [Orlov 2014]. The location of the perturbed separatrix depends not only on the perturbation field amplitude and phase, but also on the internal structure of the plasma equilibrium. The separatrix location at any given toroidal angle can be calculated in MAFOT [Wingen 2009] and the displacement amplitudes can be compared to the GPI estimates to check whether the non-axisymmetric perturbations from the RWM coils locally displace the edge flux surface as expected [Orlov 2014].

The location of the peak GPI signal as a function of time has been found via a process similar to that used to produce the top plot in Figure 6. We compare two time slices, one 5 ms before the onset of the perturbation and the other one 10 ms after the application of the perturbation. For each time slice, we use the appropriate coil current and g-file as inputs to MAFOT. An example of the current magnitudes are found in **table 5**. Using MAFOT, we predict the radial location of the outermost manifold (either stable or unstable) at a fixed Z (either midplane or GPI midpoint view) and take this as the "separatrix" viewed by GPI. To eliminate variation between shots that stem from plasma shaping, only the relative radial shift in this location is compared. Figure 11 shows the result of the analysis of 21 NSTX H-mode discharges where the radial location of the separatrix is determined via GPI before and after the application of non-axisymmetric 3D fields to be compared the with MAFOT predictions.

Table 5. Coil currents from an n=3 case used for the comparison of MAFOT and GPI.

|  | RWM1 | RWM2 | RWM3 | RWM4 | RWM5 | RWM6 |
|---|---|---|---|---|---|---|

| $I_{before}$ [A] | 0 | 0 | 0 | 0 | 0 | 0 |
| --- | --- | --- | --- | --- | --- | --- |
| $I_{after}$ [A] | 1150 | -1150 | 1150 | -1150 | 1150 | -1150 |

**Figure 11** below shows the result of the analysis of 21 NSTX H-mode discharges where the radial location of the separatrix is determined via GPI before and after the application of non-axisymmetric 3D fields to be compared the with MAFOT predictions.

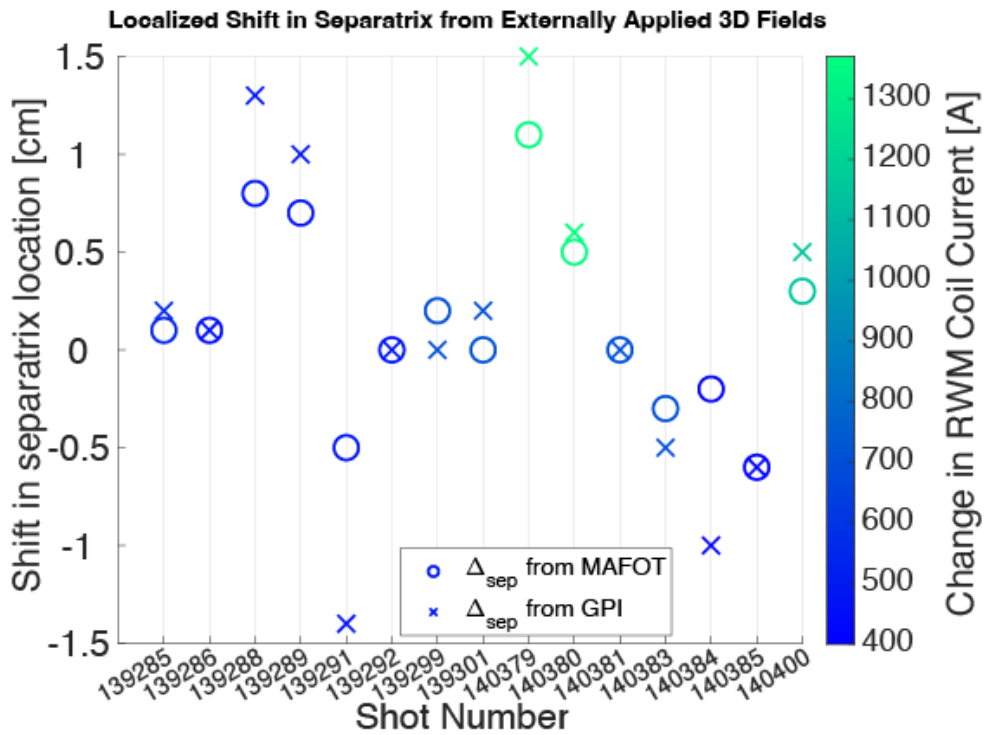

**Figure 11.** Comparison of measured shift in separatrix by GPI (X) and shift predicted by MAFOT (O) for shots in the database. The color represents the magnitude of the change in the RWM coil current amplitudes.

In general, MAFOT predicts the the separatrix displacements observed by the GPI. When the peak in the GPI intensity signal moves radially inwards, MAFOT predicts an inward shift of the outermost manifold and vice versa. It is important to note that the magnitude of the shifts measured fall within the ability of EFIT to accurately predict the location of the separatrix. The GPI intensity peak does not exactly indicate the location of the separatrix (+/- 2cm) and can vary due to the electron density and temperature profiles. However, our interest lies in the relative shift of the location of the separatrix. If it is assumed that the GPI peak moves with the flux surfaces as the 3D fields change, that uncertainty is reduced by an order of magnitude to +/- 0.2 cm and stems from the GPI optical equipment. Because the magnitude of these shifts are within or close to the uncertainty of our tools, the correlation should be quantified and compared to the correlation of a random set of points to ensure that the relationship isn't the result of some random process. In addition, it should be noted that no correlation between coil current amplitudes (3D field strength) and separatrix shift is observed.

The significance of the correlation between the plasma boundary shift measured by GPI and the shift predicted by MAFOT can be quantified by looking at the Pearson Product-Moment Correlation (Pearson's r). Calculating the r value between these two sets gives a correlation of 0.907, which typically points to a strong association between the variables. To further understand the statistical significance of this value, we first look at the average correlation coefficient of a set of two random variables that are uniformly distributed from 0 to 1. Repeating this numerical experiment 10k times then averaging the correlations results in the distribution shown in **figure 12** with a mean of 0.004. To compare correlation magnitudes, we take the absolute value of the correlation and look at the percentage of occurrences that are similar in magnitude to our data. Of the 10k correlation coefficients calculated, 2.4% are greater than 0.5 (general cut off between medium and large strength of associations), 0.5% are greater than 0.6, 0.1% are greater than 0.7, and only one case had a correlation greater than 0.8, which had a value of 0.804. Based on this distribution, we can assert that the likelihood that these two datasets have a high correlation through a random process is extremely unlikely.

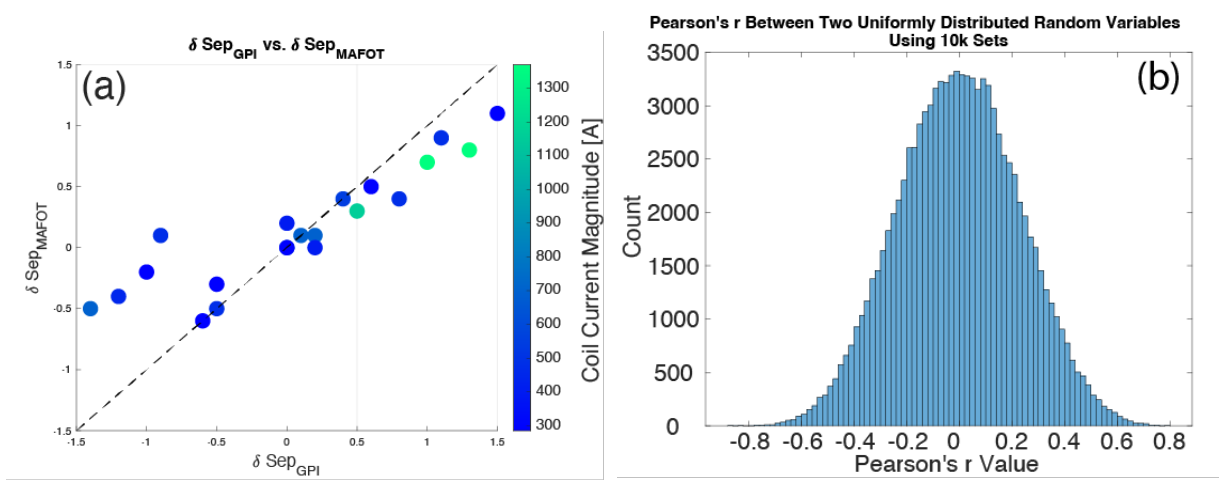

**Figure 12.** (a) Separatrix shift predicted by MAFOT vs. that measured by GPI with a strong correlation of 0.9. (b) Distribution of Pearson's r values from 10k experiments that looked at the correlation of two uniformly distributed variables of the same size of the set seen in (a).

## IV. Conclusions

In this study of the effect of externally applied 3D fields on NSTX edge turbulence, the clearest and most significant result was seen when looking at six discharges consisting of on-off and off-on 3D field switches. A weak trend is seen in Figures 7 and 8, then summarized in Tables 2 and 3, that implies that 3D fields decrease the relative fluctuation levels both inside and outside of the separatrix when in H-mode. However, these trends were only at the level of 1-2 standard deviations over a few shots, so the results did not have a high statistical significance. We were unable to find any strong statistically significant trends in either the autocorrelation time or the average poloidal size when comparing cases with and without these 3D fields.

Separatrix displacements due to changes in the externally applied 3D fields were calculated in MAFOT and compared to the experimental measurements from the GPI diagnostic on NSTX. A high correlation between the measured and predicted separatrix shifts was found. In general, the non-axisymmetric fields from the RWM coils locally displace the edge flux surface as expected even when the self-consistent plasma response is not included, and that the model reproduces the localized shift observed by GPI.

## V.     Acknowledgments

This work is supported by the US DOE Contracts DE-AC02-09CH11466 and DE-FG02-05ER54809. The authors would like to thank Dr. Galina Avdeeva and Dr. Mate Lampert for their help with NSTX signal processing.

## VI.    Appendix A

NSTX has different coordinate systems defined between physics and engineering operations, as well as that coded in MAFOT. Because of the lack of symmetry of the perturbations and the localized GPI measurement, it is important that we understand the coordinate systems being used before proceeding. To reduce confusion, we use the engineering coordinate system to map the coil names and toroidal positions to the MAFOT convention. **Figure 13** shows the two conventions used for this study.

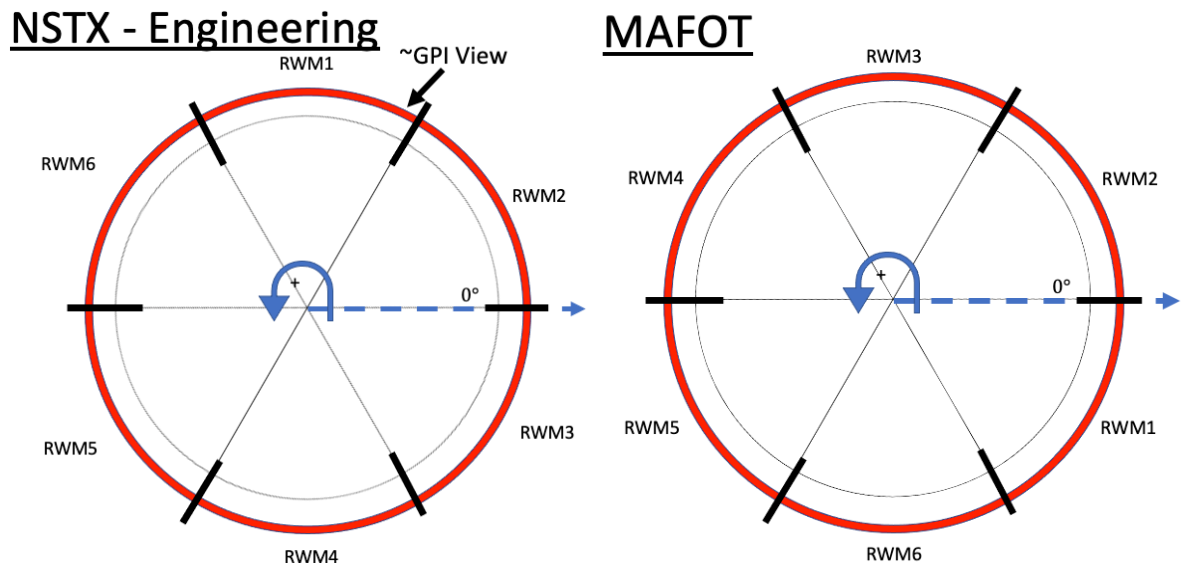

**Figure 13.** Two coordinate systems used for NSTX are shown. The engineering coordinate system on the left has it's zero between RWM2 and RWM3 and the coil number increases CW. The MAFOT coordinate system's zero is between RWM1 and RWM2 and the coil number increases CCW.

## VII.   Appendix B

For MAFOT verification, NSTX data is used to perform two numerical experiments. The first consists of synthetically rotating a perturbation around NSTX and measuring the radial location of the outermost (stable/unstable) flux surface (at a fixed z coordinate) to see if the calculated surface displacement has a sinusoidal waveform. The second numerical experiment has a fixed perturbation and instead the toroidal viewing angle is rotated (as set by MAFOT) to see if an expected sinusoidal waveform is produced. Both of these simulations use the same plasma equilibrium file and RWM coil currents from a reference discharge (NSTX shot 140394 at 563 ms) in our database.

For the rotating perturbation case, we fix the toroidal viewing angle at 0 degrees and rotate the perturbation in 60 degree increments (as set by NSTX RWM coil geometry). If the applied perturbation is purely n=1, it should result in an approximately n=1 waveform in the separatrix location.

For the fixed perturbation experiment, we fix the 270 degree perturbation from reference shot 140394 and rotate the toroidal viewing angle by 10 degrees calculating the manifold displacements in MAFOT to determine the radial location of the outermost flux surface. Again, for a purely n=1 perturbation, it is expected to produce an n=1 waveform in the radial location of the separatrix.

Results of the two numerical experiments are shown in **Figure 14** below. **Figure 14a** shows the radial position of the outermost flux surface (measured at the midplane) of the MAFOT simulation where the perturbation is fixed and the toroidal viewing angle is rotated in 10 degree increments. **Figure 14b** shows the case where the toroidal viewing angle is fixed and instead, the perturbation is rotated in 60 degree increments. After accounting for the differences in coordinate system definitions, we can overlay the two experiments to compare the predictions. As seen in **Figure 15,** there is good agreement between these two simulations. The maximum difference is seen between the 120 degree points and is ~0.1mm, which is within the resolution of the method used to determine the radial location.

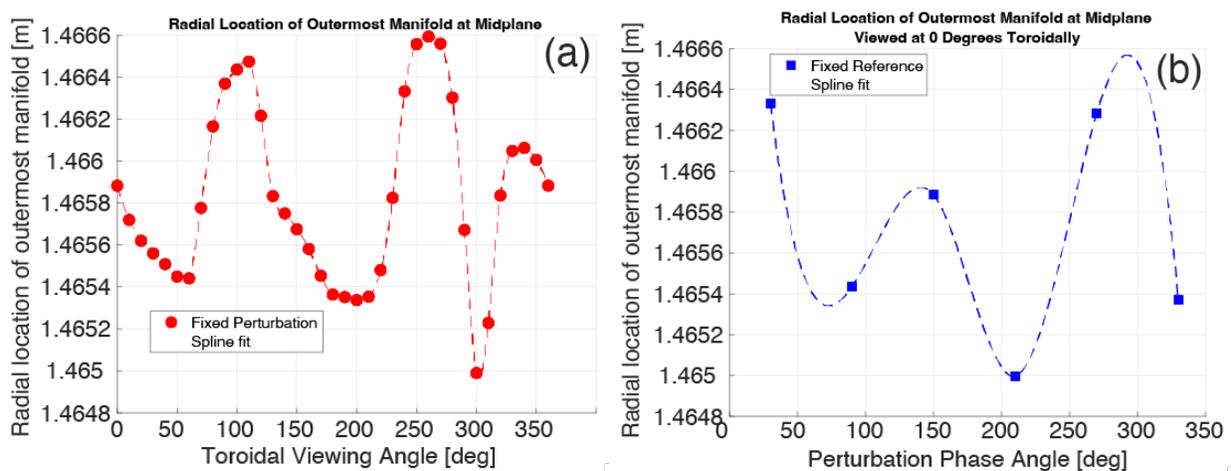

**Figure 14.** (a) Separatrix radial location as determined by MAFOT plotted vs. toroidal viewing angle for a fixed perturbation. A spline fitting is shown to illustrate the potential waveform seen

in the separatrix. (b) Separatrix radial location as determined by MAFOT plotted vs. perturbation phase angle for a fixed toroidal viewing location.

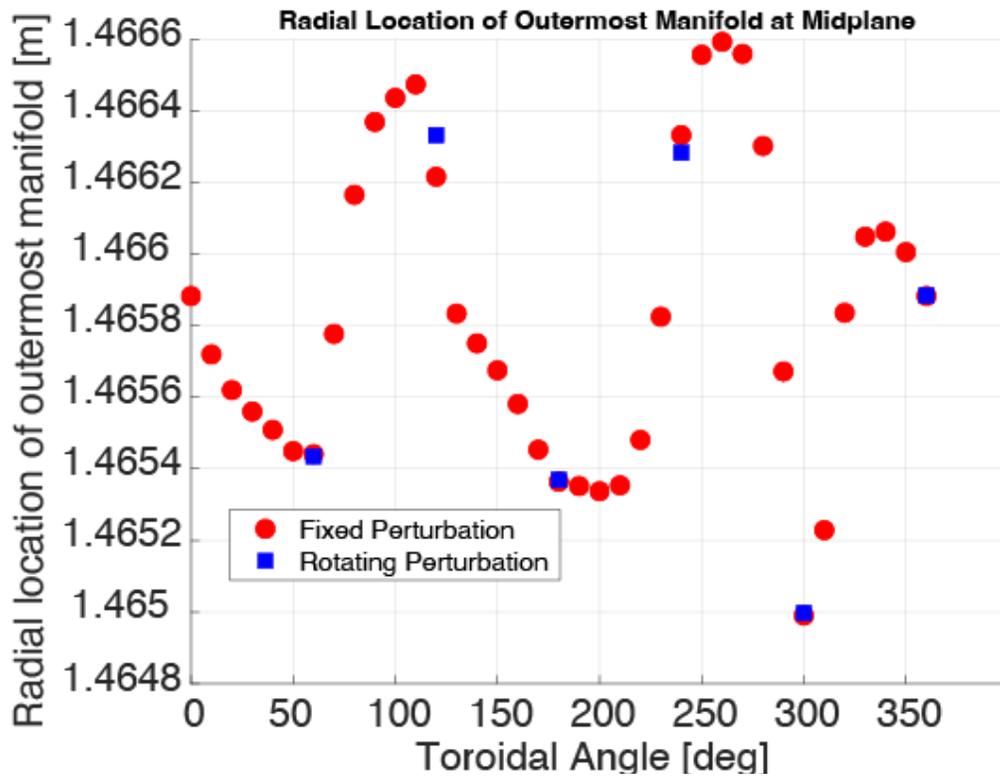

**Figure 15.** Overlay of the two numerical experiments performed with MAFOT. Red points are the radial position of the separatrix as the toroidal viewing location is varied during a fixed perturbation from the NSTX RWM coils. Blue points are separatrix radial positions measured at a fixed (0 degree) toroidal location as a perturbation is rotated around the device.